\documentclass[12pt]{iopart}
\usepackage{epsfig}  
\begin{document}

\title[Spin dynamics around a quantum critical point in Ce$_{0.87}$La$_{0.13}$Ru$_{2}$Si$_{2}$]{Pressure dependence of the spin dynamics around a quantum 
critical point :
An inelastic neutron scattering study of 
Ce$_{0.87}$La$_{0.13}$Ru$_{2}$Si$_{2}$.}

\author{S. Raymond\dag
, L.P. Regnault\dag,J. Flouquet\dag, A. Wildes\ddag 
and P. Lejay\S  
}

\address{\dag\ CEA-Grenoble, DRFMC / SPSMS / MDN, 38054 Grenoble 
Cedex, France}

\address{\ddag\ Institut Laue Langevin, 38042 Grenoble Cedex, France}

\address{\S\ CRTBT-CNRS, 38042 Grenoble Cedex, France}

\begin{abstract}

Inelastic neutron scattering experiments performed on a single crystal of the antiferromagnetic compound 
Ce$_{0.87}$La$_{0.13}$Ru$_{2}$Si$_{2}$ under applied pressures of up 
to 5 kbar
are reported.
A quantum critical point is reached at around 2.6 kbar where 
long-range magnetic order disappears. The  
variation of the characteristic energy scales with respect to
temperature and pressure is followed and found to saturate in the 
ordered phase.

\end{abstract}



\maketitle

\section{Introduction}

Much experimental and theoretical work has been 
devoted to the study of quantum phase transitions in recent years 
\cite{proced,Millis,Moriya,Sachdev}. Such a 
transition from an ordered to a disordered state occurs at zero 
temperature 
as a function of a control parameter $r$ (pressure 
$P$, 
magnetic field $H$, 
impurity concentration $x$). Heavy fermion (HF) compounds provide 
an 
opportunity to study such 
phenomena since a variety of ground states, from weak 
antiferromagnetic 
to Pauli paramagnetic states, can be realized by tuning $r$. HF 
physics has been long
 understood from the point of view of 
the so-called Doniach phase diagram describing the competition between the 
formation of a non-magnetic Kondo singlet and the realization of an 
ordered state 
via the RKKY (Ruderman-Kittel-Kasuya-Yosida) interactions \cite{Doniach}. 
Renewed interest in this field has came about from 
precise studies of 
the quantum critical point (QCP) in itinerant magnets and the marginal behavior observed near 
the QCP,
in particular  
the so-called 
non-Fermi liquid (NFL) behavior \cite{proced,Sachdev}.

Only a few studies have been  
performed on single crystals using inelastic neutron scattering (INS). The study of a single crystalline 
sample on 
a three-axis spectrometer (TAS) allows the measurement of the full 
({\bf 
Q},$\omega$) dependence of the imaginary part of the  
dynamical spin susceptibility $\chi''({\bf Q},\omega)$. At present, 
only two 
systems were extensively
studied near the QCP : CeCu$_{6}$ doped with Au on the Cu site 
\cite{Lohn} and CeRu$_{2}$Si$_{2}$ doped 
with La on the Ce site\cite{Raymond1}. 
For the latter systems experiments with Rh substitution on the Ru site
were also performed \cite{Tabata}. In these works, 
 the magnetic excitation spectrum of two samples were studied, one in the 
paramagnetic phase ($x < x_{c}$) and one located near the instability 
point 
($x=x_{c}$). While these two studies bear similar experimental 
results : reduction of the energy 
scale near the QCP and increase of the correlation length, emphasise 
was 
put on different points. For CeCu$_{6-x}$Au$_{x}$, the anisotropy of 
the 
magnetic response \cite{stok} was put forward to explain the NFL 
behavior observed 
in bulk measurements and $\omega/T$ scaling was found in the dynamical 
spin susceptibility \cite{schro}. In the latter 
Ce$_{1-x}$La$_{x}$Ru$_{2}$Si$_{2}$ system, the 
accent was put on the Self Consistent Renormalized Spin Fluctuation 
(SF) theory of Moriya \cite{Moriya} which allows to link the magnetic 
excitation spectrum 
to the bulk measurements and their evolution towards the QCP 
\cite{Raymond2,Kambe}.

Here we propose a different experimental approach 
starting from an antiferromagnetic compound and studying the 
evolution of the 
magnetic excitation spectrum with pressure up to 5 kbar. This  
spans the phase diagram through the QCP at finite temperature. The 
main advantage is 
that the same crystal is used throughout, avoiding the problem of 
disorder which is difficult to handle when results obtained on crystals 
with 
different concentration $x$ are compared. The 
disadvantage is that the temperature range is limited by the experimental 
setup (pressure cell) both on the lower 
and higher temperature sides.

\section{Experimental details}

CeRu$_{2}$Si$_{2}$ crystallizes in the body centred tetragonal 
$I4/mmm$ space group with the lattice parameters $a=b=4.197 \AA$ and 
$c=9.797 \AA$. The dependence of these parameters on La 
concentration is roughly linear and of the order of 5.10$^{-4}$ 
$\AA$/at. $\%$ La. The crystal studied here with $x$=0.13, grown by 
the 
Czochralsky method, has a volume of 250 mm$^{3}$ 
\cite{Lejay}. 

Experiments were carried out on the cold TAS IN14 
at the ILL high flux reactor, Grenoble. A first set of measurements 
was performed in a standard orange cryostat at ambient pressure and a 
second 
set was carried out in an helium transmitting medium 
pressure cell 
made of Al alloy in a large He flow orange cryostat. The 
experimental conditions were the same for both experiments using the 
constant final energy mode with 
k$_{F}$=1.97 \AA$^{-1}$. The collimations were open-40'-60'-60' and a 
graphite filter 
was used in order to reduce higher order contamination.
With this setup the width of the incoherent peak (Full Width at Half 
Maximum (FWHM) of a gaussian profile) was 0.35 meV. These 
conditions were chosen to minimize the background 
(flat analyzer and collimations) which appears to be crucial for the 
experiment in 
the pressure cell. A window 
of Cadmium (neutron absorber) was put around the cryostat in order to cut the 
background of the pressure cell. Measurements were performed
above 2.6 K to avoid the presence of superfluid He from the flow 
around the pressure cell.
\begin{figure}[h]
	\centering
	\epsfig{file=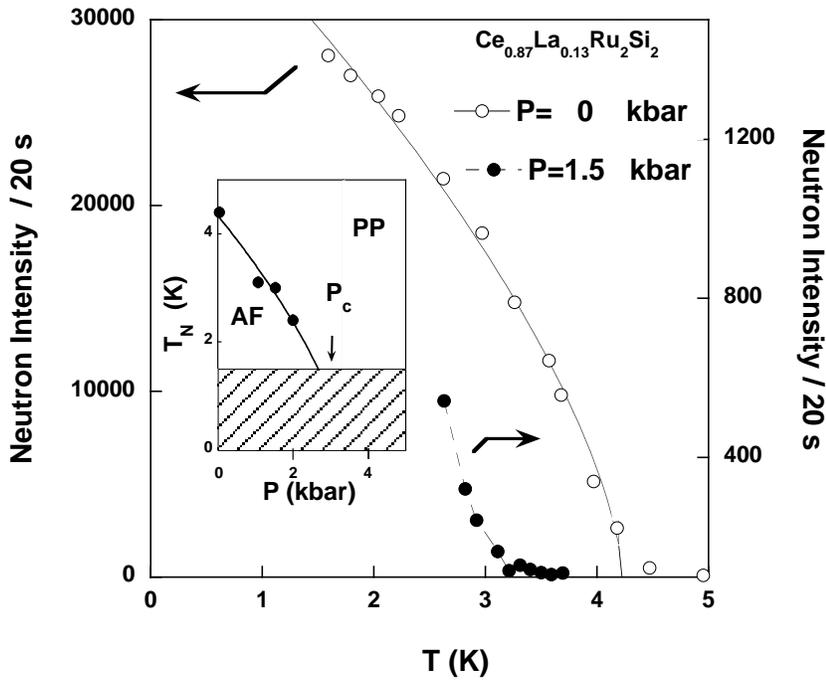,height=90mm,angle=0}
	\caption{Neutron intensity versus temperature measured at 
	{\bf Q}=(0.69,1,0) at $P$=0 and 1.5 kbar. The lines are guides for 
the eyes. 
	The inset shows the pressure variation of the N\'eel temperature. 
The 
	line corresponds to the power law fit explained in the text. The 
	hatched area corresponds to the temperature range not coverd by this 
	experimental setup.}
	\label{raymond1}
\end{figure}

\section{Determination of the critical pressure}

Magnetic ordering occurs in Ce$_{1-x}$La$_{x}$Ru$_{2}$Si$_{2}$ for 
$x$ $\geq$ 0.08 ($x_{c}$=0.075). The ordering takes the form of a sine-wave modulated 
structure with 
the incommensurate wave vector {\bf k}=(0.31,0,0) and the magnetic 
moments 
along the $c$ axis \cite{Quezel}. For the sample studied here  
($x$=0.13), the N\'eel temperature is $T_{N}$=4.4 K with a magnetic 
moment 
of 1.05 $\mu_{B}$ at 1.5 K and at ambiant 
pressure. The ($P,T$) phase diagram of this sample was 
drawn by studying the evolution of the 
magnetic Bragg peak {\bf Q}=(0.69,1,0) as a function of pressure ({\bf Q}={\bf 
$\tau$}-{\bf k} where {\bf $\tau$}=(1,1,0) is a reciprocal lattice 
translation.). The measured 
neutron intensity  
versus temperature at
ambient pressure without the cell and at 1.5 kbar in the pressure 
cell is shown in Fig.\ref{raymond1}. The pressure variation of the 
N\'eel temperature is shown in the inset of Fig.\ref{raymond1}. The 
data below 2.6 
K came from a
preliminary experiment performed with similar experimental conditions 
at 
Laboratoire  L\'eon brillouin, Saclay, France. The agreement between 
several sets of data give the experimental 
precision between two different series of experiments. The overall 
data show the 
reliability of the pressure cell. Above 2.8 kbar, no sign of 
magnetic 
ordering is found down to 1.8 K. If the pressure variation of the 
N\'eel temperature is fitted 
according to the relation expected for a 3 dimensionnal antiferromagnet near the 
QCP :
\begin{equation}
T_{N} \propto (P-P_{c})^{2/3},
\label{toto1}
\end{equation}
the critical pressure is $P_{c}$=3.4 kbar. It is also predicted that 
the $T$=0 staggered magnetic moment $m_{k}$ will follow the relation :
\begin{equation}
m_{k} \propto (P-P_{c})^{1/2}
\label{toto2}
\end{equation}
From our measurements, it is clear that $m_{k}$ does not follow a simple law 
like (2). This quantity rapidly drops at low pressure and then has 
the tendency to saturate. Such a behavior 
was preciselly studied for alloys of the family 
Ce(Ru$_{1-x}$Rh$_{x}$)$_{2}$Si$_{2}$ \cite{kawa1,kawa2} (This is not the purpose of the present 
work which concentrates on the dynamics). It is thus difficult to extract $P_{c}$ 
from the pressure variation of $m_{k}$, an upper estimate would be 2.2 
kbar. Another estimate of $P_{c}$ can 
be 
given from the bulk 
measurements performed on several concentrations or under pressure in 
this 
family of compounds. Resitivity measurements performed 
for $x$=0.2 under pressure 
show that 2.4 $\%$ of dilution corresponds to 1 kbar \cite{Haen1}. 
For the compound studied here with $x$=0.13, this 
will give an expected 
$P_{c}$
of 2.1 kbar. In the following, we will take the mean value of these 
three different estimates : $P_{c}$=2.6 
$\pm$ 0.5 kbar.

\section{Magnetic excitation spectrum at $P$=0}

\begin{figure}[h]
	\centering
	\epsfig{file=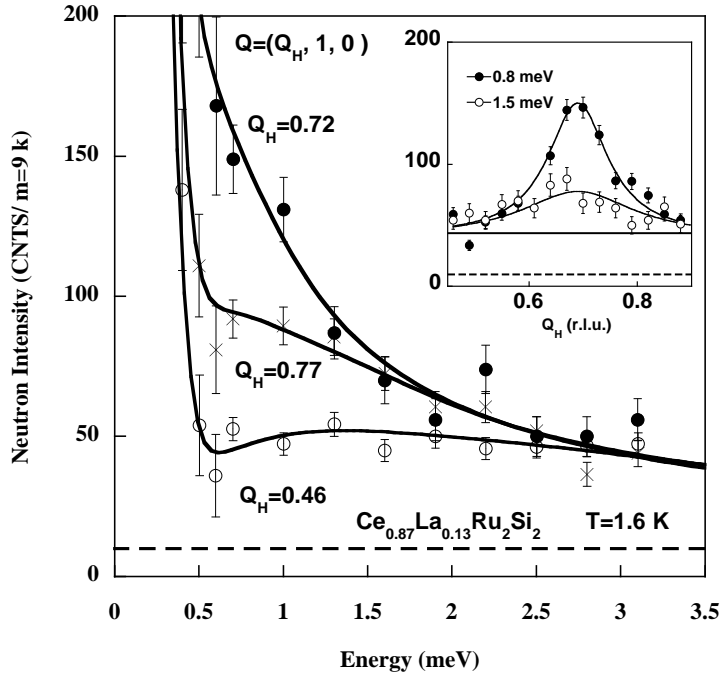,height=90mm,angle=0}
	\caption{ Constant {\bf Q} scans realized at 2.6 K and $P$=0 kbar 
	around the wavevector of instability {\bf k}. Lines corresponds to 
	the fit explained in the text. The inset shows two constant energy 
scans performed at 0.8 
	and 1.5 meV. The dashed line indicates the background of the 
	spectrometer. The solid straight line in the inset corresponds to the single site 
	contribution.}
	\label{raymond2}
\end{figure}

The magnetic excitation spectrum is shown in Fig.\ref{raymond2} for 
several 
{\bf Q} vectors ({\bf Q}=($Q_{H}$,1,0) where $Q_{H}$ is expressed in 
reciprocal lattice units (r.l.u.)) at 2.6 K. The corresponding
wave-vector response is shown in the inset of Fig.\ref{raymond2} for 
two energy transfers of 0.8 
and 1.5 meV.
The signal is peaked at $Q_{H}$=0.69 for both energies 
and 
the lineshape broadens when energy increases. The response in energy 
is quasielastic as shown in Fig.\ref{raymond2} for $Q_{H}$=0.72. This 
wave-vector is chosen to be {\bf k}+(0.03,0,0) where {\bf k} is the wave-vector of instability in order to avoid 
strong Bragg 
contamination. This offset has no consequences on the 
determined physical quantities (energy widths) since the signal we 
measured does not vary 
rapidly around {\bf k} in the energy range probed with our 
instrumental resolution. Despite the fact that the compound is 
ordered, the spectrum is 
characteristic of SF and no well defined (single mode) excitations 
are observed.
The background of the spectrometer (determined for negative energy 
transfer at low temperature) does not correspond to the 
background of the constant $\omega$ scan suggesting the 
existence 
of a {\bf Q} independent (or single-site) 
contribution as in other compounds of this familly 
\cite{Raymond2,Jaccoud,Rossat}. Consequently, we used 
 the same procedure as in the 
previous works on $x$=0 \cite{Rossat} and $x_{c}$=0.075 
\cite{Raymond2} for a quantitative analysis of the data. 
This procedure accounts for the 
background in a consistent way for both constant {\bf Q} and constant 
$\omega$ scans.  
In the first works on CeRu$_{2}$Si$_{2}$, the single-site 
contribution was not taken into account \cite{Regnault} since only 
high magnetic field studies unambiguously underline the existence of 
this contribution \cite{Jaccoud,Rossat}. By applying a magnetic 
field, the correlated part of the signal vanishes and only the 
{\bf Q} independent contribution persists at the same level of 
intensity that was determined at $H$=0. In a similar fashion to the applied 
external 
magnetic field case, we will also show  
that this decomposition of the magnetic scattering into single-site 
and correlated signal is very convenient for understanding experiments 
carried out under pressure. 
In this approach, the neutron intensity is written :

\begin{equation}
I({\bf Q},\omega)=I_{BG}+(1+n_{B}(\omega))(\chi
''_{SS}(\omega) + \chi''_{IS}({\bf Q},\omega))
\label{toto3}
\end{equation}
where $I_{BG}$ is the background intensity, 
$n_{B}(\omega)=1/(e^{\omega/T}-1)$ is the 
Bose factor and $\chi''_{SS}$ and $\chi''_{IS}$ are respectively the imaginary 
part of the single site and intersite 
magnetic dynamical susceptibility.
The single-site contribution is assumed to be Lorentzian, 
 reflecting local $4f$ spin relaxation :
 
\begin{equation}
\chi''_{SS}(\omega)=\chi'_{SS}\frac{\omega 
\Gamma_{SS}}{\omega^{2}+\Gamma_{SS}^{2}}
\label{toto4}
\end{equation} 
where $\chi'_{SS}$ is the local susceptibility and $\Gamma_{SS}$ is 
the local fluctuation rate. It is related to the Kondo temperature 
 $T_{K}$ ($\Gamma_{SS} \approx k_{B}T_{K}$ \cite{kondo}).
For the correlated signal, we use the following formula were {\bf 
q=Q-$\tau$-k} :

\begin{equation}
\chi''_{IS}({\bf q},\omega)=\frac{\chi'_{IS}({\bf q})}{2}\omega(\frac{
\Gamma_{IS}({\bf q})}{(\omega-\omega_{0}({\bf 
q}))^{2}+\Gamma_{IS}^{2}({\bf q})}+\frac{ 
\Gamma_{IS}({\bf q})}{(\omega+\omega_{0}({\bf 
q}))^{2}+\Gamma_{IS}^{2}({\bf q})})
\label{toto5}
\end{equation} 
where $\chi'_{IS}({\bf q})$ is the {\bf q} dependent part of the 
suceptibility, $\Gamma_{IS}({\bf q})$ is the intersite fluctuation rate  and 
$\omega_{0}({\bf q})$ is an inelastic energy which clearly better describes the 
data obtained for compounds of this familly located in the 
paramagnetic region 
\cite{Rossat}. In this paper, we will focus on the response in energy 
at the wave vector 
{\bf k} and will note $\Gamma_{IS} = \Gamma_{IS}(q=0)$. It is also 
found that $\omega_{0}({\bf q})$ does not depends of {\bf q} as was already known 
from studies of other compositions.

The single site contribution is unambiguously determined at {\bf 
Q}=(0.46,1,0) in a part of the Brillouin zone where 
the signal is flat in {\bf q} (see inset of Fig.\ref{raymond2}) and 
thus $\chi'_{IS}$=0. The fit to the data is shown in 
Fig.\ref{raymond2}. Since the neutron intensity is not normalized, we 
are 
only interested in the energy width, $\Gamma_{SS}$=1.4 (2) meV. 
A fit at the vector {\bf k} is also shown in Fig.\ref{raymond2}.  We 
found for the offset $q_{H}$=0.03 (${\bf q}=(q_{H},q_{K},q_{L})$) at 2.6 K in the ordered phase,  
$\Gamma_{IS}$=0.15 (3)  meV and 
$\omega_{0}$=0.10 (5) meV. Contrarily to the pure compound 
CeRu$_{2}$Si$_{2}$, the determination of $\omega_{0}$ lies in the 
limits of the fit since it is smaller than the resolution. It is 
worth noting that the determined value of $\Gamma_{IS}$ does not 
change 
much if $\omega_{0}$ is fixed to zero. For comparison, the values 
obtained at 2.6 K for the compounds corresponding to $x$=0 and 
$x_{c}$=0.075 are 
shown in Table 1. The $q_{H}$ dependence was described by expanding 
\begin{table}
\caption{Characteristic energies (in meV) measured at 2.6 K on 
several 
compounds Ce$_{1-x}$La$_{x}$Ru$_{2}$Si$_{2}$ for {\bf Q}={\bf k} with 
similar experimental conditions (cold TAS).} 
\begin{indented}
\item[]\begin{tabular}{@{}llll}
\br
x&$\Gamma_{IS}$&$\omega_{0}$&$\Gamma_{SS}$\\
\mr
0&0.77 (5)&0.5 (1)&2.0 (1)\\
0.075&0.17 (2)&0.2 (1)&1.4 (1)\\
0.13&0.15(3)&0.10(5)&1.4 (2)\\
\br
\end{tabular}
\end{indented}
\end{table}
$\Gamma_{IS}(q_{H})$=$\Gamma_{IS}$(1+$(q_{H}/\kappa)^{2}$) in (5) 
and assuming that $\chi_{IS}(q_{H})\Gamma_{IS}(q_{H})$=constant 
\cite{kura}. The fit to the 
data obtained at 0.8 and 1.5 meV are shown in the inset of 
Fig.\ref{raymond2} with a value of $\kappa$=0.05 r.l.u. This 
corresponds to a 
correlation length $\xi$ $\approx$ 13 $\AA$ that is around 3 lattice units. 
This is the correlation length of the remaining longitudinal 
fluctuations after magnetic order is established in this system. This 
must
not to be confused with 
the correlation length introduced in phase transition theory which 
diverges at $T_{N}$ or at $P_{c}$. This 
latter quantity needs to be measured in a double-axis configuration 
in order to have 
an energy integrated signal.
\begin{figure}[h]
	\centering
	\epsfig{file=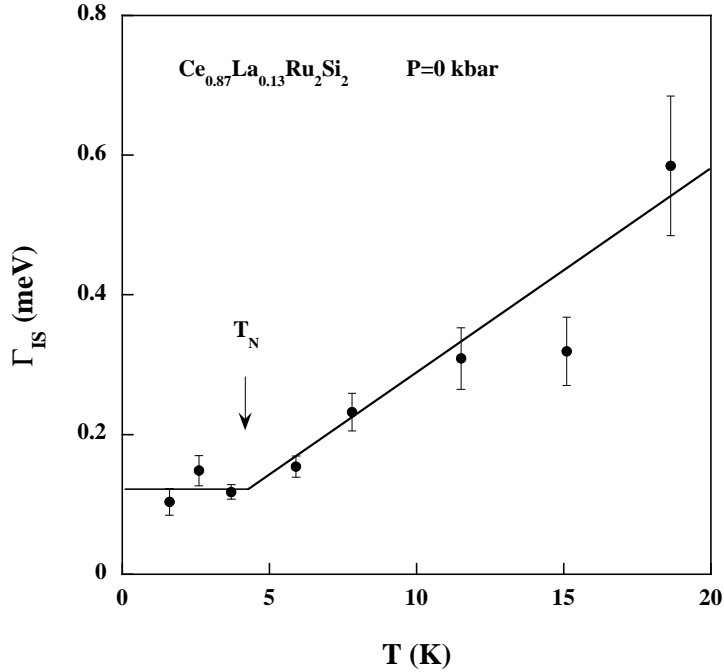,height=90mm,angle=0}
	\caption{Temperature dependence of the fluctuation rate 
$\Gamma_{IS}$ 
	measured at $P$=0 kbar. The line is a guide for the eyes.}
	\label{raymond3}
\end{figure}

The analysis was repeated for several temperatures. On 
increasing $T$, the magnetic excitation spectrum broadens 
continuously. Of interest is 
the temperature variation of $\Gamma_{IS}$. This quantity is 
shown in Fig. \ref{raymond3}. The temperature variation is almost 
linear 
above $T_{N}$, slowing down and eventualy saturating below 
$T_{N}$. In contrast, the quantity 
$\Gamma_{SS}$ is almost temperature independent in the studied range 
of 1.5-20 K. At higher temperature, for $\Gamma_{SS} > k_{B}T$, it 
is  
expected that $\Gamma_{SS}$ will acquire some temperature dependence 
\cite{kondo}.

\section{Evolution of the magnetic excitation spectrum with pressure}

\begin{figure}[h]
	\centering
	\epsfig{file=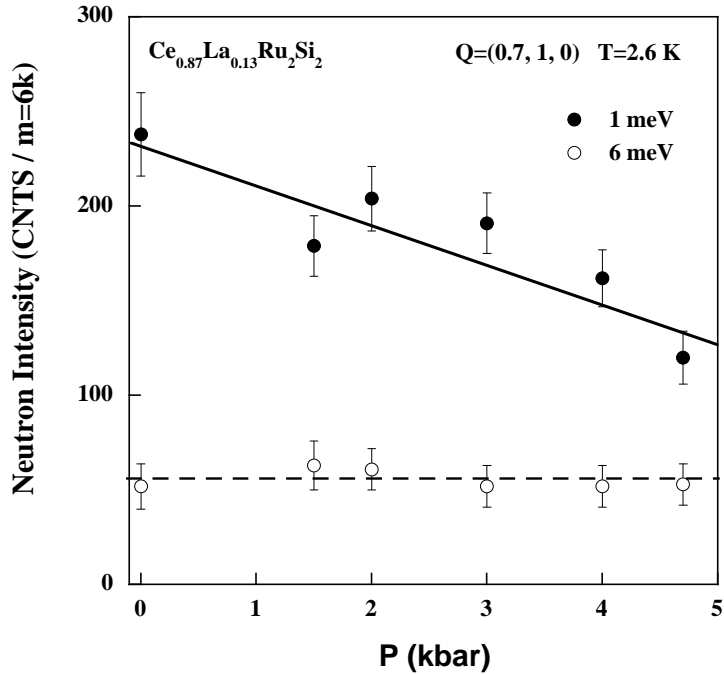,height=90mm,angle=0}
	\caption{Pressure dependence of the inelastic magnetic scattering 
	measured at {\bf Q}=(0.7,1,0) for 1 meV (full circle) and 6 meV 
	(open circle)  energy transfer at 2.6 K. 
	Spectrometer background is subtracted. Lines are guides for the 
eyes.}
	\label{raymond4}
\end{figure}

Similarly to the temperature dependence studied at zero pressure, we 
also studied the evolution of the spin dynamics 
with pressure at constant temperature of 2.6 K. Fig.\ref{raymond4} 
shows the neutron intensity measured versus pressure for two energy 
transfers of 1 and 6 meV. The spectrometer background measured for an 
energy transfer of -1 meV at 2.6 K is subtracted. 
These data show that 
the low energy response decreases when pressure increases while the 
high energy part is not sensitive to pressure. A crossover between low 
energy and high energy dynamics therefore needs to be defined. In 
the 
framework of quantum phase transition and for a gapless system 
\cite{Sachdev}, this crossover energy
$\Delta$ is the relevant quantity to study the QCP (i.e. $\Delta 
\rightarrow 0$ at the QCP). The 
analysis of the dynamical spin susceptibility of this system in a 
single-site and inter-site contributions provides 
us with a natural way of defining such a crossover. Indeed, the high 
energy part corresponds to a single-site only contribution. In the 
following, we 
will simply take $\Delta \approx \Gamma_{IS}$. In this respect 
both pressure and magnetic field suppress the inter-site contribution 
in a similar ways (See Fig.8 of Ref \cite{Rossat} in comparison with 
Fig.\ref{raymond4}). 
\begin{figure}[h]
	\centering
	\epsfig{file=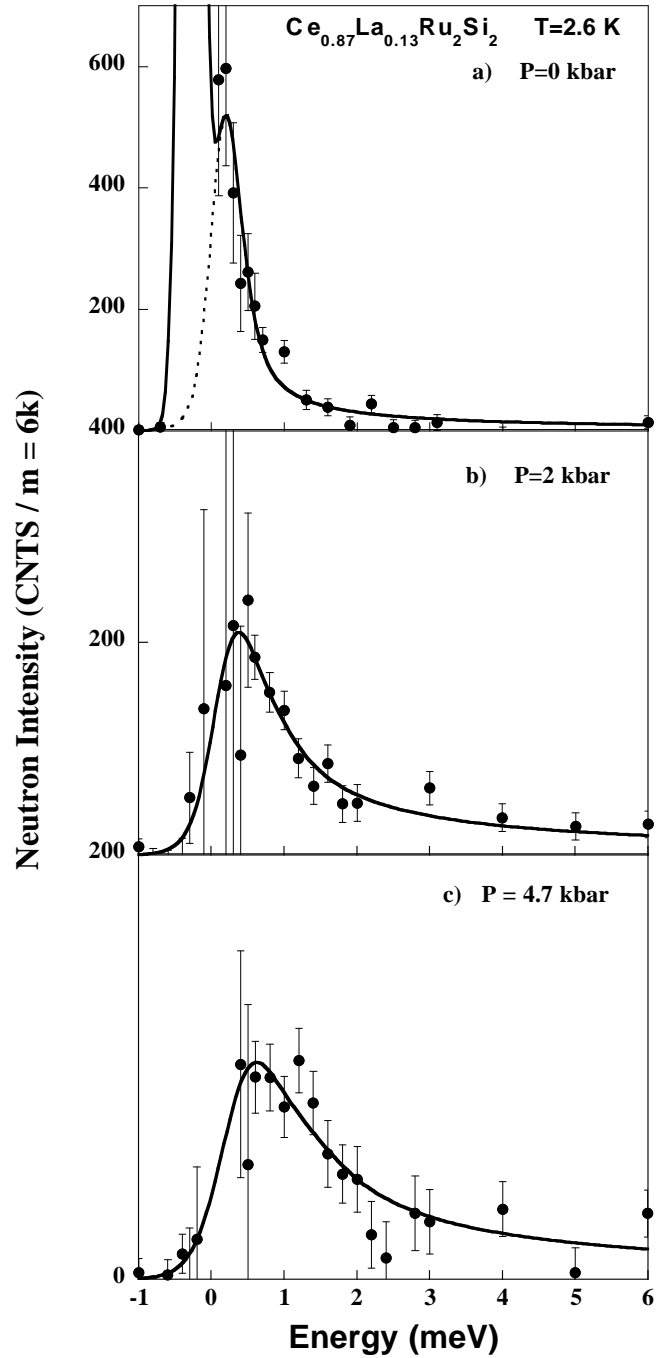,height=180mm,angle=0}
	\caption{Evolution with pressure of the intersite contribution of 
the 
	dynamical spin susceptibility measured at 2.6 K for 0, 2 and 4.7 
kbar. 
	The experimental data are obtaied by point by point subtraction of 
	{\bf Q}=(0.7,1,0)-{\bf Q}=(0.46,1,0)}
	\label{raymond5}
\end{figure}

Fig.\ref{raymond5} shows the evolution of the inter-site signal 
measured at {\bf Q}=(0.7,1,0) after subtraction of the 
data from the single-site signal measured at each pressure at {\bf 
Q}=(0.46,1,0). At zero 
pressure (upper frame), a Bragg contribution appears at negative 
energy 
transfer due to the choice of the focusing conditions (offset 
$q_{H}$=0.01). The inelastic 
signal broadens with increasing pressure and at 4.7 kbar it is very 
similar to 
the one of the pure compound CeRu$_{2}$Si$_{2}$ (see Fig.9 in Ref. 
\cite{Rossat}).

\begin{figure}[h]
	\centering
	\epsfig{file=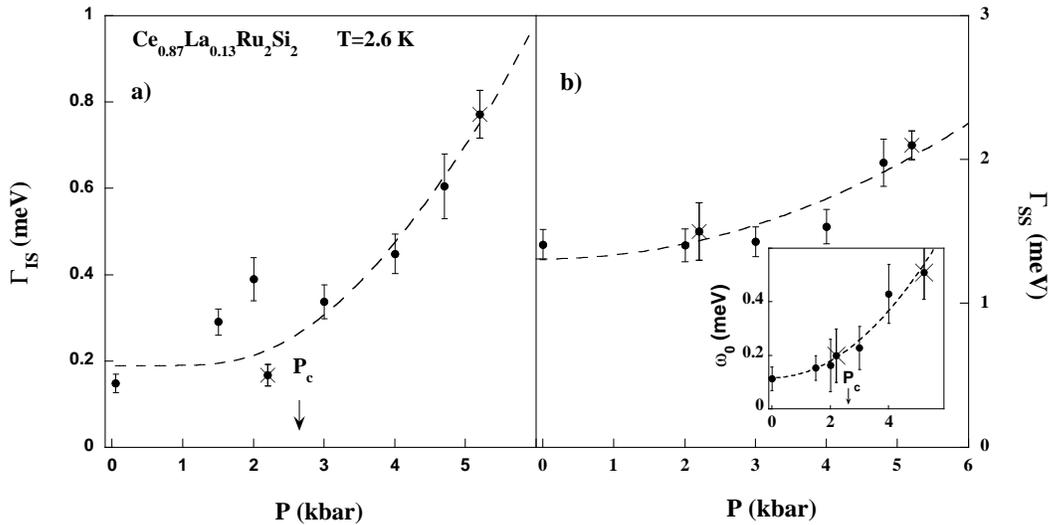,height=70mm,angle=0}
	\caption{a) Pressure dependence of the fluctuation rate 
$\Gamma_{IS}$ 
	measured at 2.6 K. b) Pressure dependence of the local fluctuation 
rate $\Gamma_{SS}$ 
	measured at 2.6 K. The inset shows the pressure dependence of 
$\omega_{0}$ . For 
	each panel, the crossed circles correspond to data obtained 
	on the alloys $x$=0 and $x_{c}$=0.075 with the pressure-concentration 
	conversion explained in the text. Lines correspond to fit explained 
	in the text.}
	\label{raymond6}
\end{figure}

The pressure variation of $\Gamma_{IS}$ is shown in 
Fig.\ref{raymond6}a. 
The data obtained at the same temperature for the alloy corresponding 
to the critical 
concentration, $x_{c}$=0.075, and to the pure compound, $x$=0, are shown 
in the same plot with 
the concentration-pressure conversion explained in the first section. 
There is an acceptable agreement with these former measurements and our 
present 
measurements under pressure. The data presented in this way exhibit some noise due 
to the fact that they were taken from different experiments with 
different setups. There is thus no correlations 
between the error bars for the points obtained under pressure and for 
the ones obtained on the alloys. 
All the more, the data taken under pressure lead also to higher error bars since the 
background is higher.
Nevertheless, this correct agreement stresses the validity of the 
analogy 
between concentration and pressure in the limit of small disorder.

Despite the limited statistics of the data, it seems that 
 the energy width saturates below the critical pressure $P_{c}$ 
as it 
does below $T_{N}$ at $P$=0. Indeed the two sets of data 
$\Gamma_{IS}(P)$ and $\Gamma_{IS}(T)$ are 
strikingly similar. The data clearly show that the 
increase of $\Gamma_{IS}$ is much higher in the non-magnetic 
phase 
above 3 kbar. To reproduce this behavior, we made a global fit of 
these data to the phenomenological expression :
\begin{equation}
\Gamma_{IS}(P)=\Gamma_{IS}(0)+\alpha P \exp(-P^{*}/P)
\label{toto6}
\end{equation} 
with $\Gamma_{IS}(0)$=0.19 (2) meV, $\alpha$=0.4 (2) meVkbar$^{-1}$ 
and $P^{*}$= 7 (2) kbar. This expression reproduces the saturation 
of $\Gamma_{IS}$ at low pressure and is linear in $P$ for $P \gg 
P^{*}$. 
The pressure variation of $\Gamma_{SS}$ is shown in 
Fig.\ref{raymond6}b. This quantity increases only slightly in the 
pressure range studied. This confirms the idea that high energy 
dynamics do not change much with pressure. The data are 
phenomenologicaly described by 
the smooth variation $\Gamma_{SS}=\Gamma_{SS}(0)+\eta P^{2}$ with 
$\Gamma_{SS}(0)$=1.3 (1) meV and $\eta$=0.025 (5)
meVkbar$^{-2}$. Finally,   
the pressure variation of $\omega_{0}$ is shown in the inset of Fig. 
\ref{raymond6}b. It is very similar to the one of $\Gamma_{IS}$.
The relation $\omega_{0}$ $\propto$ 0.6$\Gamma_{IS}$  approximately 
holds. Beyond these phenomenological descriptions, an order of 
magnitude can be extracted for the pressure varition of the different 
characteristic energies $\epsilon$ ($\epsilon=\Gamma_{IS}, 
\Gamma_{SS}$ or $\omega_{0}$). When taken the pressure range studied 
$\Delta P$ as a whole, all these quantities increase with a quite similar 
rate of $\Delta \epsilon/ \Delta P \approx$ 1 Kkbar$^{-1}$.
\section{Discussion}
\subsection{Nature of the excitations}

We first discuss the nature of the excitations in the ordered phase. 
Due to the strong Ising nature of the system \cite{Haen2}, the 
observation of 
spin waves by INS is precluded. The observed
excitations correspond to fluctuations of the order 
parameter (longitudinal fluctuations). This was previously established by INS on 
CeRu$_{2}$Si$_{2}$ \cite{jacthese}. The fact that the excitations
are not well-defined dispersive modes is understood from the magnetic 
structure of 
the system. For an incommensurate structure, {\bf q} is not a good 
quantum 
number due to the lack of translational invariance.  INS measurements 
probe the dynamical susceptibility at 
({\bf q},$\omega$) and this corresponds here to the coupling of 
several 
eigen-modes giving a broad signal \cite{Ziman}. This is not the case 
for real antiferromagnetic structure where longitudinal well-defined 
dispersive 
modes can be observed in compounds with weak magnetic moments. Such 
modes were measured by INS in 
URu$_{2}$Si$_{2}$ \cite{Broholm} . 

Our data show similar results when 
the ordered phase is reached either by varying temperature or 
pressure. The fluctuation rate seems to reach a constant and finite 
value in the ordered 
phase.  This behavior can 
be partly understood from the point of view of the magnetic sum rule which 
states that the total magnetic 
scattering integrated over {\bf q} and $\omega$
is proportional to the square of the magnetic moment of the ion (in 
fact $S(S+1)$ in the quantum mechanical treatment of a spin $S$). Since the 
ordered moment at $T$ $\approx$ 0, $P$ $\approx$ 0 ($m_{k}$=1.05 $\mu_{B}$) 
does 
not reach the saturated value ($m_{sat}$) determined by the 
crystal field ground state ($\approx$ 1.7 $\mu_{B}$), and since 
spin-waves 
are not present, the presence of longitudinal fluctuations is 
necessary in order to satisfy the magnetic sum rule. In particular, we do not observe 
any complete softening either of 
$\Gamma_{IS}(P)$ at $P_{c}$ or of $\Gamma_{IS}(T)$ at $T_{N}$. This can also be 
partly understood with the same 
argument. This is also linked to the itinerant 
nature of the magnetic order, as discussed 
in the following subsection.

\subsection{Analysis in a spin fluctuation approach}

Our data suggest that the magnetic 
order has the form of a spin density wave at least down to 
the lowest measured temperature of 1.5 K. Firstly, we are 
able to 
accurately follow the local fluctuation rate, 
$\Gamma_{SS}$ as a function of pressure, which does not vanish at the QCP but smoothly 
decreases from the disordered to the ordered state. This quantity 
reflects the mechanism of local relaxation of 4$f$ moments that is 
the Kondo 
effect. This implies that there is no breakdown of the 
Kondo effect at $P_{c}$ in the Ce$_{1-x}$La$_{x}$Ru$_{2}$Si$_{2}$ 
system. Secondly, the 
magnetic order realized for $x$=0.13 is purely sinusoidal since no 
higher harmonics were 
found down to the lowest temperatures \cite{Raymond3}. This also points 
towards a SDW 
picture since such a pure sinusoidal modulation is typical of 
itinerant 
magnetism while squaring of the modulation occurs in localized spin 
systems. Indeed, in Ce$_{1-x}$La$_{x}$Ru$_{2}$Si$_{2}$, higher 
harmonics 
were found to develop for $x \geq $ 0.2 \cite{Mignot}. Such a 
crossover from a SDW to a local 
magnetism is also observed in the similar 
Ce(Ru$_{1-x}$Rh$_{x}$)$_{2}$Si$_{2}$ compounds \cite{Kawa}. The 
underlying 
hypothesis of an itinerant magnetism description is that a 
Fermi surface is well defined throughout the whole phase diagram.  

In the past years, neutron scattering data and bulk measurements 
obtained on the Ce$_{1-x}$La$_{x}$Ru$_{2}$Si$_{2}$ system were 
self consistently analyzed in the Moriya's SF theory 
\cite{Raymond2,Kambe}. 
The finite fluctuation rate measured by neutron scattering around 
$P_{c}$ must be obviously related to the observation that the 
resistivity always 
reaches a $T^{2}$ dependence at the lowest temperature and that 
concomitently, the specific heat is always linear in $T$. Our new INS 
data 
confirm such a picture. The 
dynamical spin susceptibility, which is a phenomenological starting 
point of SF theory, 
was also recently deduced form a microscopic model 
taking equally into account both the Kondo effect and the RKKY 
interactions ; the INS cross section derived is very similar to the 
one used in this paper \cite{Lavagna1,Lavagna2}.

\subsection{Comparison with CeCu$_{6-x}$Au$_{x}$}

In the CeCu$_{6-x}$Au$_{x}$ system, the QCP is reached for $x 
\approx  0.1$. NFL 
behavior observed either at $P_{c}$ or $x_{c}$ in this system (linear 
resistivity, 
logarithmic divergence of the specific heat) was extensively studied 
by the 
Karlsruhe group \cite{Lohn}. It is believed that such a behavior 
implies a new theoretical treatement of the QCP in a strong coupling 
approach, a local picture where the Fermi liquid description breaks down 
\cite{Coleman}. The underlying idea is that both $T_{N}$ and $T_{K}$ go to zero at the QCP : the ordered phase is 
characteristic of local magnetism. 
Concerning INS, the landmark of such a behavior is the so-called 
$\omega/T$ scaling of the 
dynamical spin susceptibility \cite{schro}. 
In the description developped here, this means that 
$\Gamma_{IS}$ equals $k_{B}T$ (no single-site signal was identified in 
these studies \cite{stok,schro}.).
This obviously implies that   
(i) The fluctuation rate totally softens at the QCP\footnote{This 
point is difficult to be experimentally adressed by INS not only 
because of the limited access to the lowest temperatures but 
also because of the instrumental resolution, which will 
limit the distinction between static and dynamics below a certain 
energy $\omega$.} and (ii)
$\omega$ and $T$ are similarly weighted in the spin dynamics.
On the contrary, in the SF approach, the fluctuation rate $\Gamma_{IS}$ is written $y_{0}+a'T^{3/2}$ 
\cite{Sachdev,Lavagna2,mimi} where $y_{0} \rightarrow 0$ at the QCP and $a'$ 
is constant. Following this argument, $\omega/T^{3/2}$ scaling would be expected 
for a 3 dimensional
system near the QCP (in the cas $a'$ $\approx$ 1). This 
difference with the observed $\omega/T$ scaling in 
CeCu$_{5.9}$Au$_{0.1}$ has a deeper meaning. In the 
itinerant magnetism model \cite{Millis}, a system in the vicinity of a 
QCP is above the critical 
dimension ($d_{c}$=4 above which the Landau theory is valid). This is 
linked to the increasing importance of the  
fluctuations in time at $T$=0 near the QCP. Formaly, this is described by 
the dynamical exponent $z$ ($z$=2 for an itinerant 
antiferromagnet) and the effective dimension of the system becomes 
$d_{eff}=d+z$ where $d$ is the geometrical dimension \cite{Sachdev,Millis}. 
On the 
contrary, the scaling observed in CeCu$_{5.9}$Au$_{0.1}$ implies that 
the system is below the critical dimension. The origin of this 
behavior is believed to be in the anomalous spin dynamics 
\cite{schro} implying that the effective dimension is 
lower by 1/2 compared to the usual scenario of quantum phase 
transitions. These results need new 
theoretical treatement beyond the current understanding of the QCP.

In Ce$_{1-x}$La$_{x}$Ru$_{2}$Si$_{2}$, a complete softening of 
$\Gamma_{IS}$ is not 
observed at the QCP as already discussed. For this system, it is 
experimentally clear that $T_{N} \rightarrow 0$ at the QCP but that 
$T_{K}$
stays finite in the ordered phase. This latter quantity may collapse far in 
the magnetic phase. Concerning the temperature variation of 
$\Gamma_{IS}$, it follows the SF prediction ($\Gamma_{IS}=y_{0}+a'T^{3/2}$) with $a'$ $\approx$ 0.2 
meVK$^{-3/2}$ as measured for $x=x_{c}$. 
It is important to note that the value of $a'$ is the same for CeRu$_{2}$Si$_{2}$ and 
thus does not evolve with $P$ or $x$. The same remark holds for the 
system CeCu$_{6-x}$Au$_{x}$ : The slope of the 
order unity found de facto in the $\omega/T$ scaling for $\Gamma_{IS}$ is also 
found for pure CeCu$_{6}$ \cite{Stock2}.

The results obtained on the systems 
CeCu$_{6-x}$Au$_{x}$ and Ce$_{1-x}$La$_{x}$Ru$_{2}$Si$_{2}$ 
are quite different since the former seems to be the paradigm of 
the strong coupling theory while the latter is the paradigm of SF theory.
The behavior observed at a magnetic non-magnetic QCP is thus not universal. What still remains surprising is that pure 
CeCu$_{6}$ 
and CeRu$_{2}$Si$_{2}$  compounds are very 
similar while their respective QCP are so different. 
Regarding the low energy scales involved in 
CeCu$_{6}$, it is not excluded that the observed physics corresponds 
to the 
fact that the Fermi surface is not yet developped at the temperatures 
experimentally 
achievable. To test this possibility, $\omega/T$ scaling must be 
searched in 
Ce$_{1-x}$La$_{x}$Ru$_{2}$Si$_{2}$  for $k_{B}T \gg \Gamma_{SS}$ near 
the QCP. The idea is that such a scaling may apply in a 
temperature range above the $T^{3/2}$ regime experimentally observed and predicted by SF 
theory.

\section{Conclusion}

This study is, to our knowledge, the first INS investigation of the 
QCP 
performed on single crystal where pressure is the control parameter 
for the 
magnetic non-magnetic phase diagram. The data obtained show a 
quantitative similarity between the approach of the magnetic phase 
versus 
pressure or temperature. Our results are in agreement with the 
previous experiments performed on alloys near the QCP ($x\approx 
x_{c}$) and in the 
paramagnetic phase $x < x_{c}$. Our overall data on the compounds of 
the 
Ce$_{1-x}$La$_{x}$Ru$_{2}$Si$_{2}$ family strongly support the SF 
approach. In the future emphasise will be put on the 
low temperature data at $P_{c}$ and $x_{c}$ to confirm this picture.

\section*{Acknowledgments}

We acknowledge the help of J.M. Mignot for preliminary experiments 
performed at L.L.B., Saclay. This work benefits from usefull 
discussions on QCP with C. P\'epin, M. Lavagna, P. Haen, K. Ishida, 
K. 
Miyake, S. Kambe, B. F{\aa}k and N. Bernhoeft.

\end{document}